\documentclass[11pt]{article}
\usepackage[margin=1in]{geometry}
\usepackage{graphicx}
\usepackage{blindtext}
\usepackage{wrapfig}
\usepackage{eqnarray,amsmath}
\usepackage{xcolor}
\usepackage[utf8]{inputenc}
\usepackage[T1]{fontenc}
\usepackage{newtxtext}
\usepackage{amsthm}
\usepackage[colorlinks = true,
            linkcolor = blue,
            urlcolor  = blue,
            citecolor = blue,
            anchorcolor = blue]{hyperref}

\def\Title#1{\begin{center} {\Large #1 } \end{center}}
\def\Author#1{\begin{center}{ \sc #1} \end{center}}
\def\Address#1{\begin{center}{ \it #1} \end{center}}

\newenvironment{Abstract}{\begin{quotation}  }{\end{quotation}}
\newenvironment{Presented}{\begin{quotation} \begin{center} 
             PRESENTED AT\end{center}\bigskip 
      \begin{center}\begin{large}}{\end{large}\end{center} \end{quotation}}
\begin{document}
\begin{titlepage}
% \pubblock
\vfill
\Title{Recent Belle II results on BSM physics}
\vfill
\Author{Roberta Volpe \footnote{email: \href{mailto:roberta.volpe@cern.ch}{roberta.volpe@cern.ch}} on behalf of the Belle II Collaboration}
%\Address{INFN and Univ. Perugia, Perugia, Italy}
\Address{Department of Physics and Geology, Perugia University,\\
  Via Pascoli, 06123, Perugia, Italy}
%email: \href{mailto:me@somewhere.com}{me@somewhere.com}
\vfill
\begin{Abstract}
  We report on recent results obtained
  by the Belle II experiment related to searches
  for beyond-the-standard-model physics. The $B$ meson decay data sample used
  corresponds to an integrated luminosity of $L = 189$ fb$^{-1}$. 
  In recent years, discrepancies from standard-model predictions, which suggest violation of lepton-flavor universality,
  have emerged in multiple measurements of $B$-meson decays.  In this document two analyses aimed to search for violation
  of the light-lepton universality are reported:
  the inclusive $R(X_{e/\mu})$ measurement and an angular analysis of the $B^0 \to D^{*-} l^+ \nu_l$, with $l=e,\mu$, decays.
  In addition, beyond-the-standard-model
  physics can manifests itself with the production of new particles.
  The search for a long-lived spin-0 particle in $b\to s$ transitions, and its interpretation in two dark sector models,
  is reported. 
\end{Abstract}
\vfill
\begin{Presented}
DIS2023: XXX International Workshop on Deep-Inelastic Scattering and
Related Subjects, \\
Michigan State University, USA, 27-31 March 2023 \\
\includegraphics[width=9cm]{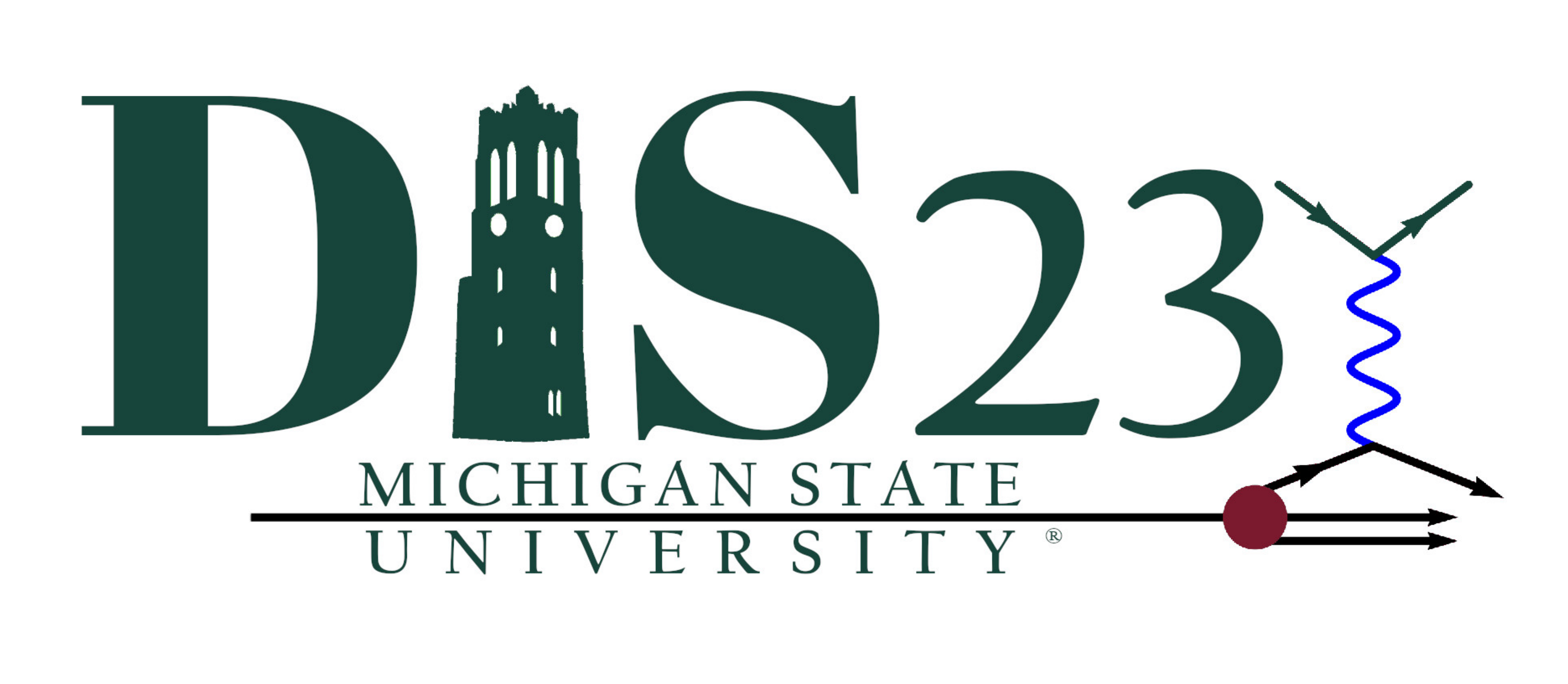}
\end{Presented}
\vfill
\end{titlepage}

\section{Introduction}
Beyond standard model (BSM) scenarios can be investigated both by searching for deviations from Standard Model (SM)
predictions in precision measurements and by searching for the production of new particles.
%If the new particles are feebly interacting with the SM particles 
Intensity frontier experiments have the capability to conduct both the searches.
The Belle II experiment is a multi-purpose detector located at the interaction point of the asymmetric $e^+ e^-$ collider SuperKEKB \cite{SuperKEKB}.
%operating
SuperKEKB mostly operates at a center-of-mass energy of $\sqrt{s}=10.58$ GeV, the mass of the $Y(4S)$ resonance,
%The $Υ(4𝑆)$
which 
decays almost exclusively into $B\bar{B}$ pairs,
making it a so-called $B$-factory.
%The high luminosity,
%producing
The production of a large amount of $B$-meson pairs,   
and the clean environment of electron-positron collisions constitute exemplary conditions to 
carry out a complete $B$-physics program and to search for feebly interacting particles
which can be part of dark-sector scenarios.
The Belle II detector is arranged around the beam pipe in a cylindrical structure,
it is composed of a hermetic magnetic spectrometer surrounded by particle-identification detectors,
an electromagnetic calorimeter, and a $K_L^0$ and muon detector.
High reconstruction efficiency and momentum resolution for charged particles and photons as well as excellent
particle identification allow searches for rare $B$ meson decays. 
The solid-angle coverage of over 90\% and well known initial state
are crucial to investigate final states with missing energy.
%High reconstruction efficiency for neutral particles and excellent resolutions for ..\\
Details of the experimental apparatus are described in \cite{belle2}.
The Belle II detector has already collected, between 2019 and 2022, a data sample corresponding to an integrated luminosity of 362 fb$^{-1}$ at the $Y(4S)$ resonance.
%collected between 2019 and 2022.
%and other samples 
This document reports on the results of two analyses aimed to test the light-lepton universality in Sec.\ref{sec:LU} 
and a search for a long-lived particle Sec.\ref{sec:DS}.
The three analyses use a data sample collected at the $Y(4S)$ resonance between 2019 and 2021,
corresponding to 189 fb$^{-1}$ of integrated luminosity.
An additional sample of 18 fb$^{-1}$, collected at an energy 60 MeV below the $Y(4S)$ mass, is used as a control sample
to estimate the $e^+ e^- \to q \bar q$ ($q=u dsc$) background and it is referred to as \textit{off-resonance} in the following.

%related to the
%a measurement of the ratio of branching fractions of inclusive semileptonic 𝐵 meson
%decays, 𝑅(𝑋 𝑒/𝜇 ) = B(𝐵 → 𝑋𝑒𝜈)/B(𝐵 → 𝑋𝜇𝜈)

%One of the two $B$ mesons in the event is reconstructed in decays to hadronic final states using the Full
%Event Interpretation package of Belle II \cite{fei}.

%high reconstruction efficiency for neutral particles and excellent resolutions 
%...\\
%Belle II experiment is

%The Standard Model (SM) 

%The Belle2 detector...\\
%Searches for BSM physics
%...\\
%Belle II is a multi-purpose experiment.

%an almost complete upgrade of the original Belle experiment, with better performance and higher rate capabilities [1]. It is located at the SuperKEKB e
%+e
%− collider at
%the KEK laboratory in Tsukuba, Japan.

\section{Tests of light-lepton universality}
\label{sec:LU}
%In the SM all leptons share the same electroweak coupling, a symmetry referred to as lepton
% universality (LU).

The gauge lagrangian of the SM predicts that the different charged leptons, the electron, muon and tau, have identical electroweak interaction strengths.
This symmetry is known as lepton universality (LU) and 
in the past years there were some hints of a violation of this symmetry.
%For example
Indeed the combination of several measurements of the ratio $R(D^{(*)}) = B(B\to D^{(*)} \tau \nu)/ B(B\to D^{(*)} l \nu)$,
with $l= e, \mu$, manifests some tension with the SM value \cite{RD}.
%\cite{RD2023}.
%( )𝑅 𝐷 (∗) = B(𝐵 → 𝐷 (∗) 𝜏𝜈)/B(𝐵 → 𝐷 (∗) ℓ𝜈), ℓ ∈ {𝑒, 𝜇)%They involved especially $\tau$ leptons %among leptons. 
%...\\
In the next sections two analyses are described where the universality between electron and muon is tested, this symmetry is named
\textit{light-lepton universality}.
A unique feature of $e^+ e^-$ colliders is the possibility to precisely estimate the energy of the center of mass interaction and
to fully reconstruct the hadronic decays of one the two $B$ mesons, referred to as $B_{\textrm{tag}}$.
This
%allows to set kinematical constraints for
kinematically constrains
the signal $B$
%and it
which is crucial to study signal $B$ decays 
with missing energy.
This \textit{hadronic $B$ tagging} is carried out by using the
full-event interpretation package of Belle II \cite{fei}.
The analyses described in this section, which study processes with neutrinos in the final state,
%take full advantage of
use this tool.
%exploit this tool to constraint  
%Both are based on the hadronic tag 
%One of the two $B$ mesons in the event is reconstructed in decays to hadronic final states using the Full
%Event Interpretation package of Belle II \cite{fei}.

\subsection{$R(X_{e/\mu})$ measurement}
\label{sec:RXem}
%ciccio
This section briefly reports 
the measurement of 
$R (X_{e/\mu}) = B(B\to X e \nu_e)/ B(B\to X \mu \nu_{\mu})$,
where $X$ indicates the generic hadronic final state of the semileptonic
$B$ decay originating from 
$b \to c l \nu$ or $b \to u l \nu$.
%The \cite{fei} 
%The FEI algorithm is used to reconstruct the
The analysis is described in detail in \cite{rx_belle2}.
After the $B_{\textrm{tag}}$ is reconstructed, the signal side is selected by requiring
%a lepton ($e$ or $\mu$)
light-lepton
candidate with a momentum, in the rest frame of the $B_{\textrm{sig}}$, $p^B_{l}$,  greater than 1.3 GeV.
%The lepton charge is requested to be opposite to the charge of the $B_{tag}$.
%\begin{figure}[htb]
  \begin{wrapfigure}{r}{0.5\textwidth}
  \centering
  \includegraphics[height=6.5cm]{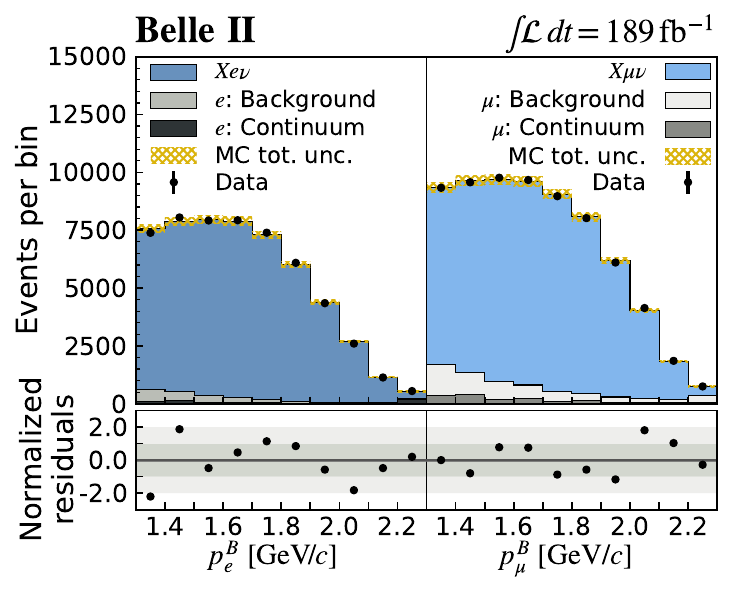}
 \caption{\small Spectra of $p_l^B$ with the fit results overlaid.}
 \label{fig:rx}
 \end{wrapfigure}
  %\end{figure}
The lepton charge is requested to be opposite to the charge of the $B_{\textrm{tag}}$.  
A boosted decision tree (BDT), built with event-topology variables, is used to reduce
continuum-background contributions.
%The lepton charge is requested to be opposite to the charge of the $B_{tag}$.
A control sample, defined with the lepton having the same charge of the $B_{\textrm{tag}}$,
is used to constraint the background contribution due to hadrons identified as leptons
or leptons originating from charmed hadrons, named \textit{background} in Figure \ref{fig:rx}.
The continuum-background contribution is estimated with the off-resonance sample.
%For each lepton $l$,
The $p^B_{e}$ and $p^B_{\mu}$ distributions,
%the $p^B_{l}$ distribution
are fitted simultaneously to extract the number of the signal events in the two flavor channels.
The main systematic uncertainties are due to the lepton identification efficiency and hadron mis-identification weights.
After correcting for efficiencies and acceptances,
%the measured $R(X_{e/\mu})$ results:
\begin{equation}
R(X_{e/\mu}) = 1.007 \pm 0.009 \text{ (stat) } \pm  0.019 \text{ (syst), }
\end{equation}
which agrees with with a previous measurement from Belle in exclusive $B \to D^* l \nu$ decays \cite{rd_belle}
and with the SM value \cite{rx_sm}.

%\newpage
\subsection{Angular asymmetries of hadronically tagged $B^0 \to D^{*-} l^+ \nu_l$ decays}
\begin{figure}[htb]
  \centering
  \includegraphics[width= 0.98\textwidth]{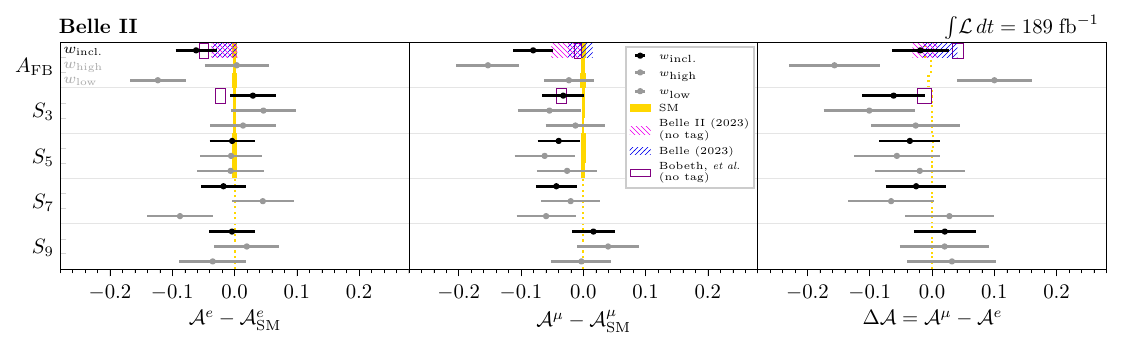}
  \caption{\small Measured asymmetries and asymmetry differences (points), one-standard-deviation bands from the previous Belle \cite{angularbelle} and Belle II
    \cite{angularbelle2}
    measurements (hatched boxes) and Bobeth et al. \cite{bobeth} (empty boxes), and SM expectations (solid boxes).}
  \label{fig:angular}
\end{figure}

%A reinterpretation \cite{bobeth} of Belle data \cite{belle}, cunducted with only one angular variable 
We report on the first comprehensive test of light-lepton universality in the angular distributions
of semileptonic $B$-meson decays to vector charmed mesons. The detailed description of the analysis is in \cite{angularbelle2}.
%The detailed description of the analysis is in \cite{angularbelle2}.
\begin{wrapfigure}{r}{0.4\textwidth}
%\begin{figure}[htb]
  \centering
  \includegraphics[width=0.36\textwidth]{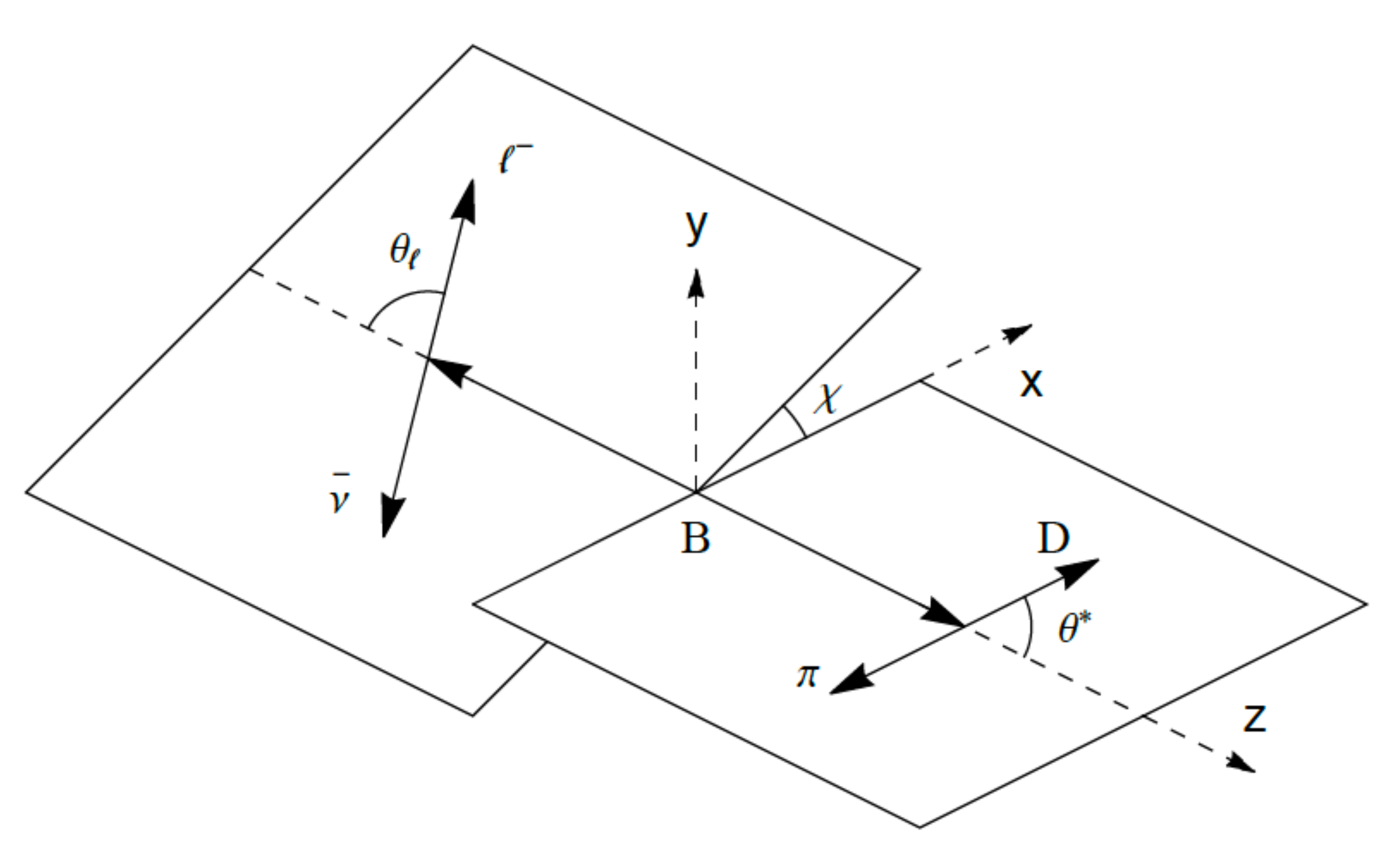}
  \caption{\small Illustration of helicity angles, from \cite{angularvar}.}
  \label{fig:helicity}
  %\end{figure}
\end{wrapfigure}
%The detailed description of the analysis is in \cite{angularbelle2}.
The five variables considered are designed to cancel most theoretical and
experimental uncertainties to be maximally sensitive
to LU violation (LUV); they are proposed and described in detail in \cite{angularvar}
and consist of one- or two-dimensional integrals of the differential decay rate as a function of the recoil parameter
$ w = \frac{m_B^2 + m_{D^*}^2 - q^2 c^2}{2 m_B m_{D^*}}$,
%where m B and m D ∗ are the B and D ∗ masses and q is the four-vector of the momentum transfer.
where $q$ is the four-vector of the momentum transfer, $m_B$ and $m_{D^*}$ are the $B$ and $D^*$ masses
and the helicity angles $\theta_l$, $\theta_V$ and $\chi$ are defined in Figure \ref{fig:helicity}.
%The five variables considered are designed to cancel most theoretical and experimental uncertainties and to be maximally sensitive
%to LU violation (LUV),
%%\cite{angularvar}.
%they are proposed and described in detail in \cite{angularvar}
%and consist of
%one- or two-dimensional integrals of the differential decay rate as a function of the recoil parameter
%$w$
%%\begin{equation}
% $ w = \frac{m_B^2 + m_{D^*}^2 - q^2 c^2}{2 m_B m_{D^*}}$
%%  \end{equation}
%and the helicity angles $\theta_l$, $\theta_V$ and $\chi$, defined in Fig.\ref{fig:elicity}.\\
The definition of the angular asymmetries, called $A_{FB}$, $S_3$, $S_5$, $S_7$ and $S_9$, is:
\begin{equation}
 \mathcal{A}_x (w) \equiv  \Big( \frac{d\Gamma}{dw}  \Big)^{-1} \Big[ \int_0^1 - \int_{-1}^0 \Big] dx  \frac{d^2 \Gamma}{dw dx}, 
\end{equation}
with $x= \cos \theta_l$ for $A_{FB}$, $\cos 2\chi$ for $S_3$, $\cos\chi \cos\theta_V$ for $S_5$, $\sin\chi \cos\theta_V$ for $S_7$ and $\sin 2\chi$ for $S_9$.
%$A_{FB}(cos \theta_l, w)$, $S_3(cos 2\chi,w)$, $S_5(cos\chi cos\theta_V,w)$,
%$S_7(sin\chi cos\theta_V, w)$ and $S_9(sin 2\chi, w)$.
%The angular asymmetries
They are evaluated by measuring the number of events $N^{-}_x$ with $x \in [-1,0]$ and $N^{+}_x$ with $x \in [0,1]$:
\begin{equation}
  \mathcal{A}_x (w) = \frac{N_x^+(w)-N_x^-(w)}{N_x^+(w)+N_x^-(w)}.
%  \text{ and }
%  \Delta \mathcal{A}_x(w) = \mathcal{A}^{\mu}_x(w) - \mathcal{A}^{e}_x(w)
\end{equation}
The light lepton universality is tested by estimating  $\Delta \mathcal{A}_x(w) = \mathcal{A}^{\mu}_x(w) - \mathcal{A}^{e}_x(w)$.
The integral is measured in three $w$ ranges: the full phase space
($w$ incl. ), $w \in $ [1.0, 1.275] ($w$ low ) and $w \in   [1.275, \sim 1.5]$ ($w$ high).
Only one $B$ tag candidate is kept based on the highest value of the algorithm’s output classifier \cite{fei}.
The signal $B$ candidate is reconstructed with  
$B^0 \to D^{*-} l^+ \nu_l$ with $ D^{*-} \to \bar{D}^0 \pi^-$ and
$\bar{D}^0 \to K^+ \pi^-$, $K^+ \pi^- \pi^+ \pi^-$, $K^+ \pi^- \pi^0$, $K^+ \pi^- \pi^+ \pi^- \pi^0$,
$K_S^0 \pi^+ \pi^-$ , $K_S^0 \pi^+ \pi^- \pi^0$, $K_S^0 \pi^0$ or $K^+ K^-$ (and charge coniugate decays).
The main selection requirements are as follows: origin of the tracks close to the interaction point, particle identification conditions,
constraints on the invariant masses for intermediate mesons decays and a missing energy larger than 0.3 GeV.
The distribution of $M^2_{miss} = (p_{Y(4S)}-p_{B_{\textrm{tag}}}-p_{D^*} - p_{l})^2$, peaking close to 0 for signal events correctly reconstructed,
is used to discriminate between signal and background events and
evaluate, by means of a binned likelihood fit, the quantities $N_x^+(w)$ and $N_x^-(w)$.
Selection efficiency, detector acceptance and bin migration 
are taken into account and corrected for.
The dominant uncertainties are statistical,
while the largest systematic uncertainty is due to the limited sample of simulated events.
Figure \ref{fig:angular} summarizes the results of 
the measurements.
%of the angular asymmetries.
$\chi^2$ tests have been performed to test the agreement with the SM expectations for different $w$ ranges
and no indication of LUV is observed.
%A reinterpretation \cite{bobeth} of Belle data \cite{belle}, cunducted with only one angular variable  

\section{Search for long lived particles in $b\to s$ transitions}
\label{sec:DS}
The search for a long-lived spin-0 particle, in $b\to s$ transition is motivated by several BSM scenarios.
In particular the new particle, named $S$ in the following, could be a mediator between the SM sector and the dark sector.
By assuming that $S$ decays exclusively to SM particles and that the coupling to SM is small, it should be long-lived.
The most
%accredited
%acknowledged
used benchmark
models foresee $S$ as a scalar coupling to the SM Higgs boson \cite{scalar} or $S$ a pseudo-scalar,
the axion-like particle (ALP) \cite{alp}.
The analysis presented in \cite{llp} searches for the decays $B^+ \to K^+ S(\to x^+ x^-)$ and
$B^0 \to K^{*0}(\to K^+ \pi^-)  S(\to x^+ x^-)$, where $x^+ x^- = e^+ e^-, \mu^+ \mu^-, \pi^+ \pi^-$ or $K^+ K^-$,
with an $S$ lifetime in the range $0.001 < c\tau <100$ cm.
The selection proceeds with a requirement of a $K^+$ or a $K^{*0} \to K^+ \pi^-$
and two tracks.
To take into account several $c\tau$ values, the selection of the two tracks requires prompt decays or a displaced vertex, the performance on the reconstruction of a displaced vertex is checked using a $K_S$ control sample.
The signal is extracted with extended maximum likelihood fits to the unbinned distribution of a modified invariant mass
$M'(x^+ x^-) = \sqrt{M^2(x^+x^-)-4m_x^2}$, that is easier to model near threshold than $M^2(x^+x^-)$.
The $M'$ distribution for the signal is described with a double Crystal Ball function and has a
%$M^2(x^+x^-)$.
resolution, $\sigma_{\textrm{sig}}$, depending on the $S$ mass $m_S$ ranging from 2 MeV$/c^2$ (for small $m_S$)
to 10 MeV$/c^2$ (for large $m_S$).
The background is parameterized with a straight line, added to an exponential function in some cases,
with normalization and slope from fit to data.
%Independent fits are performed for the eight channels and for several values of generated mass
%($ m_S \in [0.025, 4.78]$ GeV$/c^2$) and lifetime ($c\tau \in [0.001, 400]$ cm ) and
%no significative excess over the SM background has been observed,
%being the maximum global significance of $1.0 \sigma$.

\begin{wrapfigure}{r}{0.6\textwidth}
%  \begin{figure}[htb]
  \centering
   \includegraphics[height=12cm]{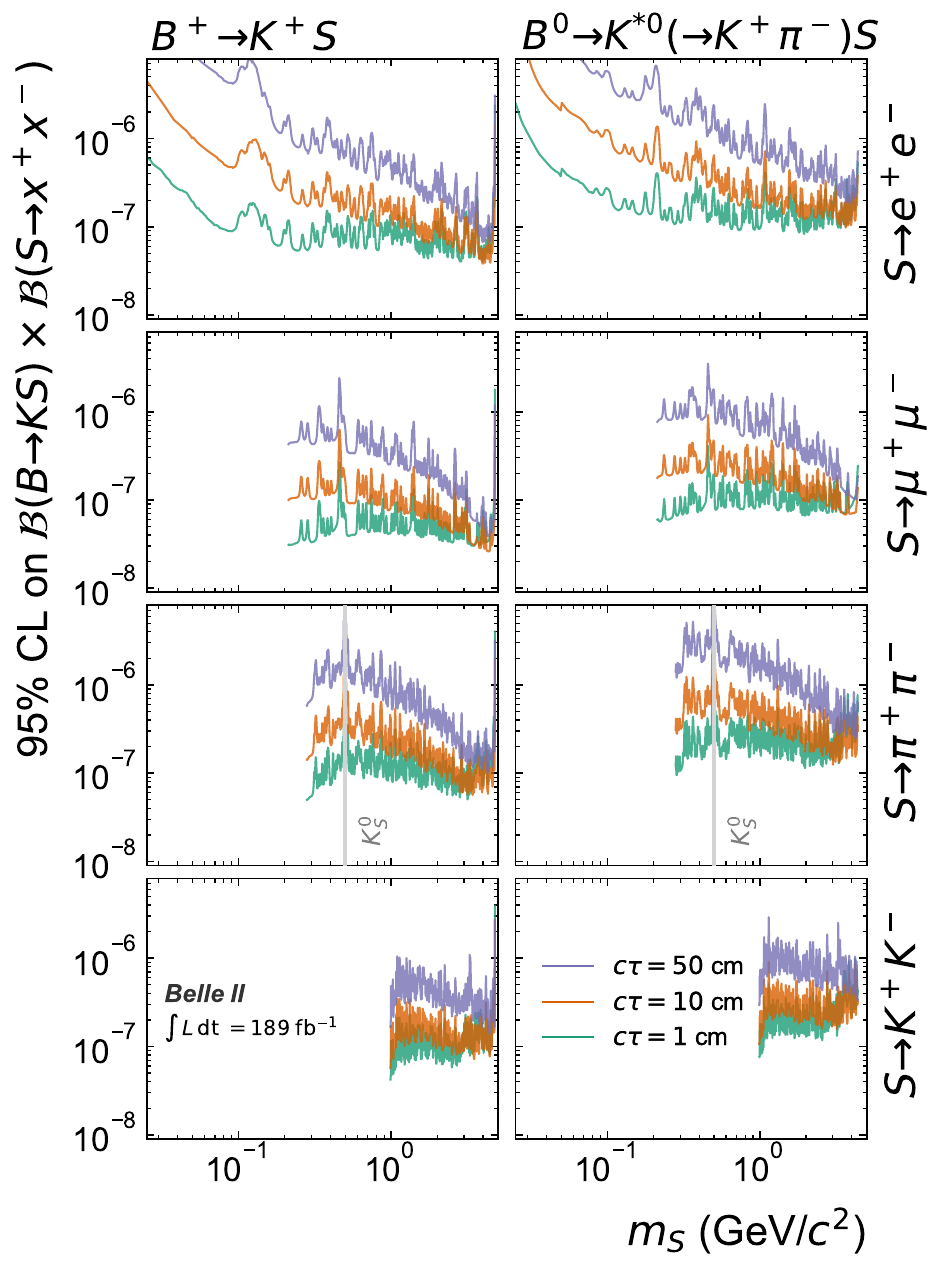}
 \caption{\small Upper limits (95\% CL) on the product of branching fractions $B(B^+ \to K^+ S ) \times B(S \to x^+ x^- )$ and  $B(B^0 \to K^*0(\to K^+ \pi^-) S ) \times B(S \to x^+ x^- )$ as functions of scalar mass $m_{S}$. The grey band indicates the vetoed $K_S \to \pi^+ \pi^-$ region.}
  \label{fig:B2KS}
%\end{figure}
\end{wrapfigure}
%
%with a global significance

\begin{figure}[htb]
 \centering
 \includegraphics[width=0.4\linewidth]{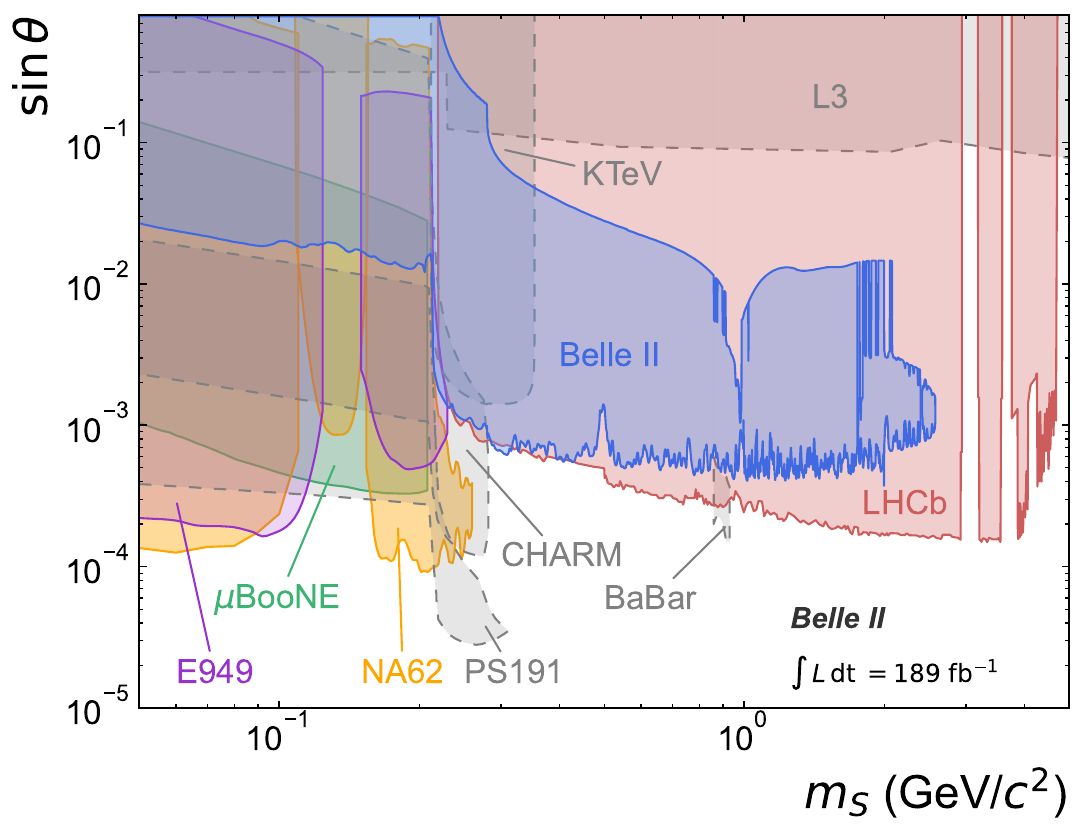}
 \hspace{0.5cm}
 \includegraphics[width=0.4\linewidth]{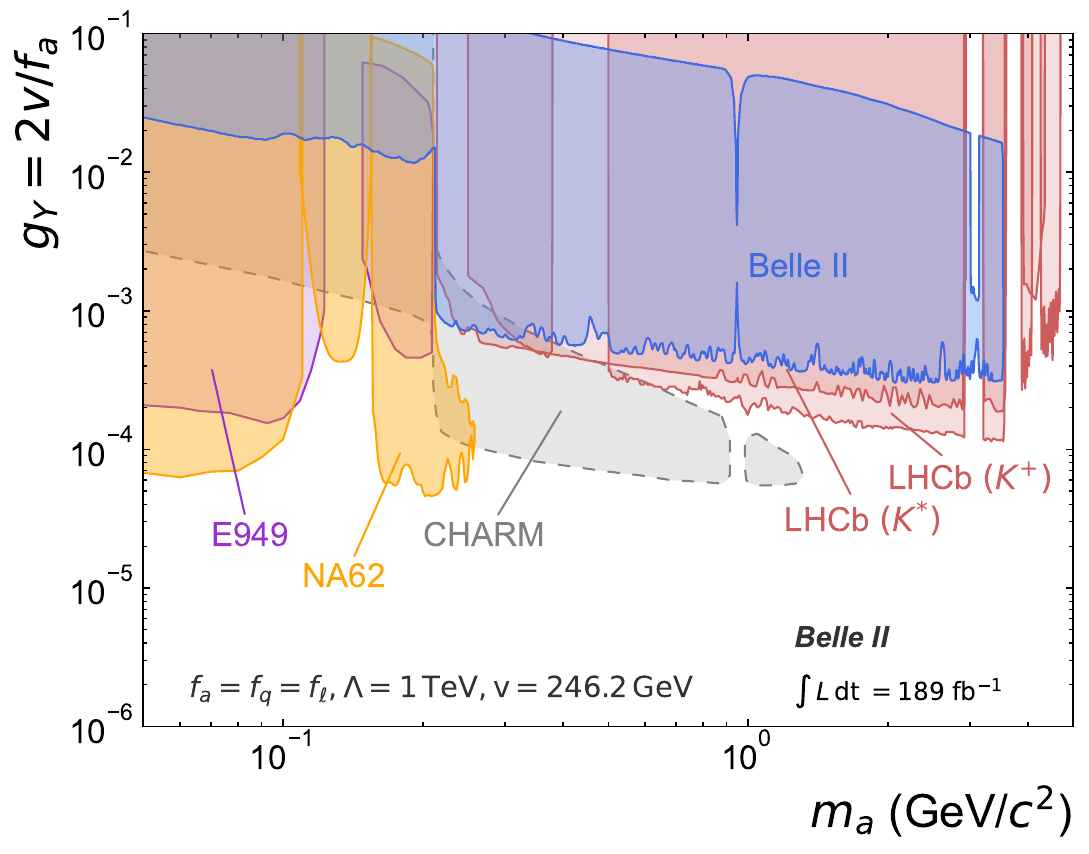}
 \caption{Exclusion regions at 95\% CL in the plane of $\sin\theta$ and $m_S$ for the dark scalar model (left) and
   $g_Y$ and $m_a$ for the ALP model (PS191, KTeV, E949, NA62, and BABAR limits correspond to 90\% CL).
 }
  \label{fig:int}
\end{figure}

Independent fits are performed for the eight channels and for several values of generated mass
($ m_S \in [0.025, 4.78]$ GeV$/c^2$) and lifetime ($c\tau \in [0.001, 400]$ cm), each fit is performed on a $M'$ range corresponding to $\pm 20 \sigma_{\textrm{sig}}$.
For $ x^+ x^- = \pi^+ \pi^-$ the $M'$ region close to the $K_S$ mass is
%fully
vetoed. 
No significative excess over the SM background has been observed;
%being
the maximum global significance is $1.0 \sigma$.
Upper limits to the product $B(B \to K^+ S ) \times B(S \to x^+ x^- )$ are set for the same mass and lifetime ranges.
%mass range $ m_S \in [0.025, 4.78]$ GeV$/c^2$.
Figure \ref{fig:B2KS} shows the 95\% confidence level (CL) observed upper limits obtained
with the $CL_S$ method with asymptotic approximation as a function
of $m_S$ for three values of $c \tau$.
%These limits result in
The strongest bounds for $S\to e^+ e^-$ has been obtained and
for the first time limits are set on $S$ decaying to hadrons.\\
%\begin{wrapfigure}{r}{0.6\textwidth}
%\begin{figure}[htb]  
%  \centering
%  \begin{tabular}{cc}
%  \includegraphics[width=0.4\linewidth]{modelindependent_B2KS_belle2.pdf} &
%  \includegraphics[width=0.35\linewidth]{scalar_summary.pdf}\\
%   & \includegraphics[width=0.35\linewidth]{alps_fermions_bc10.pdf}
%  \end{tabular}
%  \caption{\small Left: Upper limits (95\% CL) on the product of branching fractions $B(B^+ \to K^+ S ) \times B(S \to x^+ x^- )$ and
%    $B(B^0 \to K^*0(\to K^+ \pi^-) S ) \times B(S \to x^+ x^- )$ as functions of scalar mass $m_{S}$.}
%  \label{fig:B2KS}
%\end{figure}
%\end{wrapfigure}
%\begin{tabular}{cc}
%\includegraphics[width=8cm]{modelindependent_B2KS_belle2.pdf}
%\end{tabular}  
%\begin{wrapfigure}{r}{0.4\textwidth}
%  \centering
%\includegraphics[width=0.4\linewidth]{modelindependent_B2KS_belle2.pdf}
% \caption{\small Left: Upper limits (95\% CL) on the product of branching fractions $B(B^+ \to K^+ S ) \times B(S \to x^+ x^- )$ and
%    $B(B^0 \to K^*0(\to K^+ \pi^-) S ) \times B(S \to x^+ x^- )$ as functions of scalar mass $m_{S}$.}
%  \label{fig:B2KS}
%\end{figure}
%\end{wrapfigure}
The eight channels are combined to obtain the exclusion regions for the two models where $S$
is a dark scalar mixing with the Higgs
%with a coupling $sin \theta$
and an ALP coupling to photons,
fermions or gluons.
Figure \ref{fig:int} reports the 95\% CL exclusion region in the plane defined by a parameter of the model
%to the SM
versus $m_S$ in the two interpretations.\\
%Fig.\ref{fig:int} (left) reports the 95\% CL exclusion region in the plane $sin \theta$,
For the dark scalar interpretation the used parameter is $\sin\theta$, with $\theta$ 
representing the mixing angle of $S$ with the SM Higgs, 
%coupling
%to the Higgs,
versus $m_S$,
%while Fig.\ref{fig:int} (left)
while for the ALP interpretation it is $g_Y=2 v / f_a$,
with $v$ indicating the vacuum expectation value and $f_a$ indicating the coupling to the fermions, versus the mass of the ALP $m_a$.
This work provides the strongest constraint for $m_S \sim 0.3$ GeV$/c^2$.
%\newpage

\section{Summary}
This document reports three recent 
results achieved by the Belle II collaboration on BSM physics
with $L=189$ fb$^{-1}$.
%a part of the  
The two analyses testing the ligh-lepton universality give an outcome
in agreement with the SM and can be considered a preparation to the LU tests.
The search for long-lived particles $S$ results in no
significative excess over the SM background.
First limits are set on $S$ decaying to hadrons and the most stringent limits have been obtained for $S\to e^+ e^-$.

%The following page limits are recommended but not strictly enforced:
%\begin{itemize}
%\item Plenary talks: 8 pages
%\item Parallel talks: 5 pages
%\end{itemize}

\end{document}